\documentclass[letterpaper]{article} 
\usepackage{aaai2026}  
\usepackage{times}  
\usepackage{helvet}  
\usepackage{courier}  
\usepackage[hyphens]{url}  
\usepackage{graphicx} 
\usepackage{natbib}  
\usepackage{caption} 
\usepackage{algorithm}
\usepackage{algorithmic}

\usepackage{booktabs}
\usepackage{multirow}
\usepackage{amsmath}
\usepackage{tabularx}
\usepackage{xcolor}
\usepackage{newfloat}
\usepackage{listings}
\DeclareCaptionStyle{ruled}{labelfont=normalfont,labelsep=colon,strut=off} 
\floatstyle{ruled}
\newfloat{listing}{tb}{lst}{}
\floatname{listing}{Listing}
\nocopyright 

\title{MapAgent: Trajectory-Constructed Memory-Augmented Planning for Mobile Task Automation}
\author{
    Yi Kong \textsuperscript{\rm 1}, Dianxi Shi \textsuperscript{\rm 2}\thanks{Corresponding author.}, Guoli Yang \textsuperscript{\rm 2}, Zhang ke-di \textsuperscript{\rm 1}, Chenlin Huang \textsuperscript{\rm 1}, Xiaopeng Li \textsuperscript{\rm 1}, \\Songchang Jin \textsuperscript{\rm 3}\\
}
\affiliations{
    \textsuperscript{\rm 1}National University of Defense Technology, Changsha, China\\

    \textsuperscript{\rm 2}Advanced Institute of Big Data, Beijing, China\\
    \textsuperscript{\rm 3}Intelligent Game and Decision Lab, Beijing, China\\
    \{kongyyy, dxshi, zhangkedi10, clhuang, xiaopengli\}@nudt.edu.cn, \{yanggl\}@aibd.ac.cn, \{jsc04\}@tsinghua.org.cn
    
}

\usepackage{bibentry}

\begin{document}

\maketitle

\begin{abstract}
The recent advancement of autonomous agents powered by Large Language Models (LLMs) has demonstrated significant potential for automating tasks on mobile devices through graphical user interfaces (GUIs). Despite initial progress, these agents still face challenges when handling complex real-world tasks. These challenges arise from a lack of knowledge about real-life mobile applications in LLM-based agents, which may lead to ineffective task planning and even cause hallucinations. 
To address these challenges, we propose a novel LLM-based agent framework called MapAgent that leverages memory constructed from historical trajectories to augment current task planning.
Specifically, we first propose a trajectory-based memory mechanism that transforms task execution trajectories into a reusable and structured page-memory database. Each page within a trajectory is extracted as a compact yet comprehensive snapshot, capturing both its UI layout and functional context. 
Secondly, we introduce a coarse-to-fine task planning approach that retrieves relevant pages from the memory database based on similarity and injects them into the LLM planner to compensate for potential deficiencies in understanding real-world app scenarios, thereby achieving more informed and context-aware task planning. 
Finally, planned tasks are transformed into executable actions through a task executor supported by a dual-LLM architecture, ensuring effective tracking of task progress.
Experimental results in real-world scenarios demonstrate that MapAgent achieves superior performance to existing methods. The code will be open-sourced to support further research.

\end{abstract}


\section{Introduction}
In recent years, Large Language Models (LLMs) have advanced rapidly,  driven by large-scale training data. These sophisticated models \cite{achiam2023gpt, liu2024deepseek} exhibit remarkable abilities in information understanding, logical inference, and content generation across various fields \cite{hadi2023survey, zhao2023survey}. The emergence of Multimodal LLMs (MLLMs), which incorporate cross-modal perception capabilities \cite{zhang2024mm, zhang2025unified}, has effectively broken through the inherent limitations of text-only LLMs. Agents built on the foundation of these LLMs (both LLM and MLLM are referred as LLM for simplicity in this paper) have demonstrated autonomous task performance abilities in a wide range of areas \cite{Wang_2024, liu2025advanceschallengesfoundationagents}, such as software development \cite{hong2024metagpt, li2023camel}, social behavior simulation \cite{park2023generative, tang2024gensimgeneralsocialsimulation}, medical application \cite{kim2024mdagents, li2025agenthospitalsimulacrumhospital}, and gaming scenarios \cite{qin2024mp5, tan2024cradle}.

\begin{figure*}[!ht]
  \centering
  \includegraphics[
  width=1.0\textwidth,
  keepaspectratio]{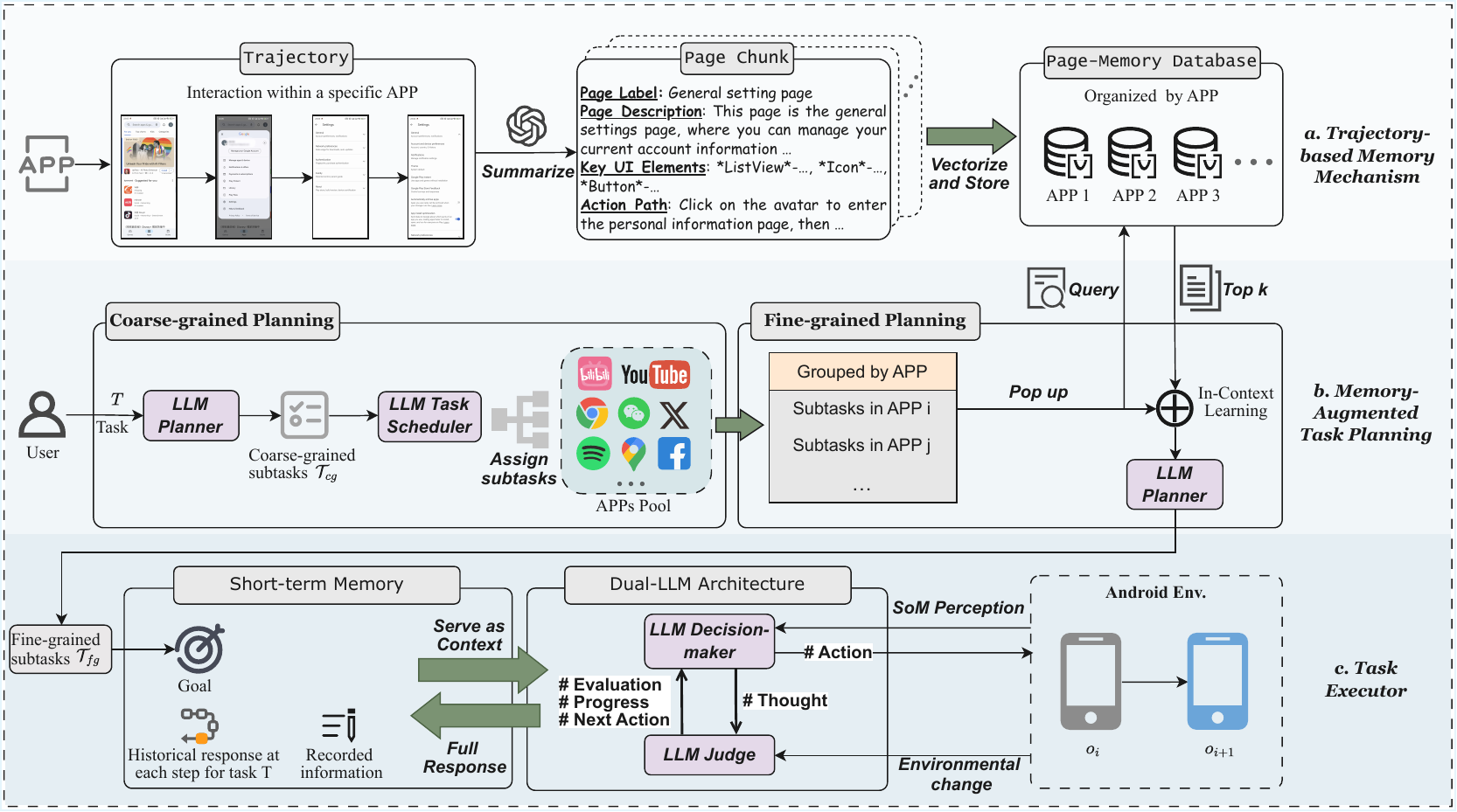}
  \caption{Overview of MapAgent framework. 
  (a) We employ a multimodal LLM to summarize each page of a trajectory within an app as a page chunk, which is subsequently vectorized and stored in the page-memory database. (b) The proposed coarse-to-fine task planning approach, initially utilizes the LLM to plan user tasks and schedule subtasks corresponding to specific apps, followed by retrieving pages from the memory database as contextual information to augment planning. (c) A dual-LLM architecture task executor equipped with short-term memory, which consists of an LLM Decision-maker and an LLM Judge that collaborate to complete planned tasks.
  }
  \vspace{-0.2cm}
  \label{fig: overview}
\end{figure*}

Among these, task automation on mobile devices stands out as a significant application. In contrast to traditional API-based virtual assistants such as Siri or Google Assistant, which are predominantly governed by rigid predefined rules, agents bolstered by these large foundation models and grounded in graphical user interfaces (GUIs) have emerged as promising solutions for automating flexible and complex user tasks \cite{zhang2025largelanguagemodelbrainedgui, liu2025llmpoweredguiagentsphone}. These LLM-based GUI agents can perform human-like interactions (e.g. click, type, scroll, etc.) on mobile devices through multimodal perception \cite{qin2025ui, gou2025uground}, complex architectures \cite{mobileagentv2, zhang2024mobileexperts}, and adaptive planning \cite{zhu2025moba, zhou2025chop}. 
This emerging operational paradigm follows a sequential execution process, where LLM undergoes reasoning and action generation for each environment observation. This represents a major advancement over traditional rule-based approaches, transitioning to complex systems capable of dynamic, context-aware operations across various mobile applications.



Although existing research has shown promising results, certain limitations persist for real-world applications \cite{liu2025llmpoweredguiagentsphone}. In practical scenarios, the agent plans the user's intention across one or more apps to accomplish a task. However, since most state-of-the-art LLMs are autoregressive models trained on extensive yet limited datasets, they cannot essentially understand the real world, resulting in potential hallucinations \cite{zhang2023sirenssongaiocean, LeiHallucination}. This issue is equally prevalent when LLM-based agents attempt to comprehend and operate various mobile apps. In contrast to LLM, humans engage in trial and error when using a mobile phone or navigating an app for the first time. Throughout this process, individuals develop mental representations of their devices and the apps they use. When faced with a specific task within an app again, they subconsciously draw upon these memories to mentally outline the entire procedure before taking action.

Inspired by the process of human subconscious task planning, we propose a novel LLM-based agent framework called MapAgent that leverages memory constructed from historical trajectories to augment current task planning. 
MapAgent consists of three key components (see Figure \ref{fig: overview}):
\textbf{a) Trajectory-based memory mechanism.} Mimicking how humans remember information during the operation process, the memory mechanism condenses historical trajectories into a structured page-memory database. Ensures that critical information encountered during execution is retained and organized.
\textbf{b) Memory-augmented task planning.} Through a coarse-to-fine planning approach, this component retrieves relevant pages from the page-memory database via similarity-based matching. It enables the agent to draw upon past experiences and stored information, facilitating more informed and context-aware planning.
\textbf{c) Task executor.} Implemented as a dual-LLM architecture execution engine, the executor is seamlessly integrated with the memory mechanism and the planning approach. Translates generated plans into concrete actions while continuously tracking task progress, thus handling the complexities and dynamics inherent in mobile environments.
To validate our method in real-world scenarios, we conduct experimental evaluations on the actual device across English and Chinese contexts.
The results of the experiment on the SPA-Bench \cite{chen2025spabench} and the CHOP dataset \cite{zhou2025chop} which both simulate daily smartphone usage scenarios show superior performance to SOTA agents.
Ablation analysis further confirms the effectiveness of our memory-augmented planning approach for mobile task automation.

In summary, our contributions can be outlined as follows:
\begin{itemize}
  \item
    We propose MapAgent framework for mobile task automation. It strategically augments the LLM agents' planning ability by storing past trajectory experiences and retrieving them based on their similarity to the current task.
  \item
  MapAgent is equipped with a dual-LLM architecture task executor, which handles dynamics and complexities in the mobile application environment through the collaborative work of dual LLM roles.
  \item
  Through comprehensive experiments in real-world scenarios, we demonstrate the superior effectiveness of MapAgent over existing methods and its promising potential in the domain of AI-assisted smartphone development.
\end{itemize}

\section{Related Work}
\subsection{LLM-based Agent}
LLM-based autonomous agents have recently emerged as a research hotspot in the field of artificial intelligence. By integrating reasoning skills \cite{CoT, yao2023react}, function calling \cite{zhuang2024toolchain, Patil2024Gorilla}, and meticulous orchestration \cite{zhuge2024gptswarm}, these agents can independently execute tasks within specific environments, demonstrating significant potential across various domains. AutoGPT \cite{auto_gpt} is a pioneering agent in this field, capable of interacting with users and decomposing LLM actions into distinct parts to fulfill user requests. MetaGPT \cite{hong2024metagpt} coordinates the collaboration of multiple agents through Standardized Operating Procedures (SOPs) to automate the software development process. The Chain-of-Discussion framework \cite{tao2024chainofdiscussionmultimodelframeworkcomplex} explores the potential to improve content generation quality by employing multiple agents engaged in debate mechanisms. Agent Hospital \cite{li2025agenthospitalsimulacrumhospital} constructs a simulated hospital environment populated by adaptive medical agents, mimicking the process of doctors diagnosing and treating patients. With the emergence of multimodal LLMs, agents have evolved from single-text modal to systems supporting multimodal interaction, leading to real-world applications such as computer use \cite{wu2024oscopilotgeneralistcomputeragents, hu2024agents} and robot control \cite{chen2024mapgpt, zawalski2025roboticcontrolembodiedchainofthought}.

\subsection{Mobile Task Automation}

Mobile task automation aims to autonomously execute user tasks through device interaction. Traditional API-based approaches require extensive development efforts and exhibit limited adaptability to dynamic interfaces \cite{zhang2025apiagentsvsgui}.
Early GUI-based methods based on supervised learning \cite{seq2act, burns2022motifvln} or reinforcement learning \cite{toyama2021androidenv, glidersigir21} are highly dependent on human demonstrations, which limits their scalability in dynamic, real-world applications. Recently, LLM-based GUI agents have made significant progress through advanced language understanding, multimodal perception, and reasoning ability \cite{zhang2025largelanguagemodelbrainedgui}. 
SeeAct \cite{zheng2024seeact} leverages GPT-4V \cite{gpt4v_system_card} for visual comprehension and employs HTML-guided action grounding to achieve task automation.
AutoDroid \cite{AutoDroid2024} uses HTML-style GUI representation and task-based memory architecture to assist in LLM decision-making. AppAgent \cite{appagent} learns to interact with specific UI elements through autonomous exploration or observing human demonstrations. Mobile-Agent-v2 \cite{mobileagentv2} explores the potential of multi-agent systems in handling complex mobile tasks. M3A \cite{rawles2025androidworld} combines ReAct \cite{yao2023react} and Reflexion \cite{ShinnCGNY23} prompting with SoM \cite{yang2023setofmark} visual annotations for zero-shot Android control and cross-app generalization. 
Despite these advancements, existing methods have not effectively addressed the knowledge bias that arises when LLMs function as planners or decision-makers in real-world applications.
Our method mitigates this issue by efficiently retrieving task-relevant page knowledge from trajectory-constructed memory and incorporating it as additional context for the agent.


\section{Preliminary}

\paragraph{Problem Setup}
Mobile task automation involves multi-step sequential decision-making.
Given a user task $T$, based on the current environment observation $o_t \in \mathcal{O}$ and previous actions $\{a_0, a_1, \dots, a_{t-1}\} \in \mathcal{A}$, the agent (acting as a policy model $\pi$) should generate an action for step $t$. This process is represented by the following formula:
\begin{equation}
    a_{t}=\pi(T,o_{t},\left \{a_{0},a_{1},\dots ,a_{t-1}\right \} )
\end{equation}
The historical trajectory $h_t$ is composed of the observation and action sequence at the time step $t$, and can be mathematically represented as follows:
\begin{equation}
    h_{t} = \left \{o_{0},a_{0},o_{1},a_{1},\dots ,a_{t-1}, o_{t} \right \}
\end{equation}
When the task is completed or exceeds the maximum steps, the whole trajectory ends.

\paragraph{Observation Space $\mathcal{O}$}
We use device screenshots and Android XML files \cite{AutoDroid2024, appagent} as the observation space. Screenshots serve as real-time visual perception information, and XML, a structured representation, is used to enhance the agent's understanding of the environment, enabling it to grasp element attributes and states.

\paragraph{Action Space $\mathcal{A}$}
To enhance the agent's ability to recognize the operable elements of the device and interact with them seamlessly, we extract the interactive elements from the XML file and use the SoM \cite{yang2023setofmark} visual prompt method to tag each element with a numerical label, assisting the agent in action grounding. Appendix A provides a detailed description of the action space.

\begin{figure*}
  \centering
  \includegraphics[width=1.0\textwidth]{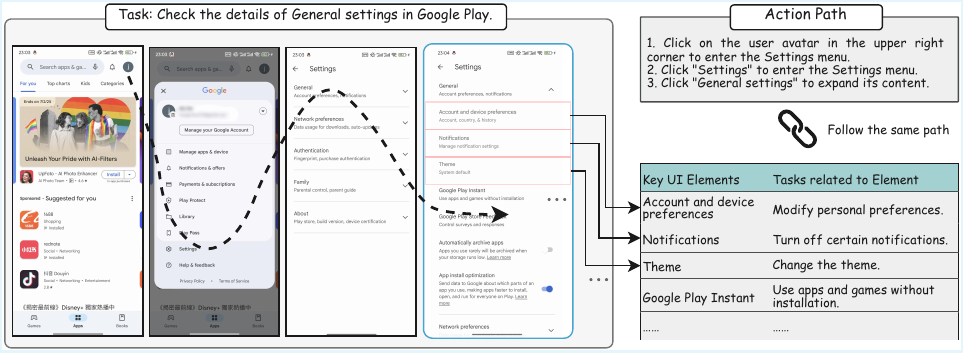}
  \caption{
  When the task \textit{Check the details of General settings in Google Play} is completed, taking the last page in this execution trajectory as an example, all tasks related to the UI Elements of this page can follow the same execution path.}
  \vspace{-0.2cm}
  \label{fig: page case}
\end{figure*}

\section{Method}
This section elaborates on the core method of the proposed MapAgent framework, which comprises three primary components, each driven by different LLM roles. 
As shown in Figure \ref{fig: overview}, the trajectory-based memory mechanism transforms pages within trajectories into structured page knowledge through an LLM summarizer and stores them in a memory database. During the planning phase, a coarse-to-fine task planning method consisting of an LLM Task Scheduler and an LLM Planner first schedules tasks to appropriate apps and then retrieves relevant pages from the memory database to augment planning. Finally, a dual-LLM architecture task executor executes the planned tasks through an iterative collaboration between an LLM Decision-maker and an LLM Judge. We will detail each component in the following subsections. 
All prompts for each LLM role are described in Appendix D.

\subsection{Trajectory-based Memory Mechanism}
Mobile task automation can be broken down into page arrival and page operation \cite{wu2025reachagent}, and the agent relies on trial and error to reach the target page. However, even though the agent has previously performed relevant tasks and successfully reached the target page, it still needs to undergo the same trial-and-error process when navigating this page again. This behavior is analogous to a person using a mobile phone or app but failing to retain any memory of its contents. To address this limitation, we propose a trajectory-based memory mechanism that distills information from trajectories to enhance the effectiveness of future interactions. 

When the agent performs a task, an operation triggers a page transition, thereby forming a trajectory consisting of a sequence of actions and pages. The inspiration for memory mechanism is that a trajectory is not only relevant to the original task, but can also extend to tasks involving other UI elements present on each page. For instance, as illustrated in Figure \ref{fig: page case}, the key UI elements of these pages in the task trajectory can serve as \textit{subconscious memory}, akin to how humans mentally retain information during the operation process. Furthermore, the action path of this task trajectory can be applied to tasks related to these UI elements.

Specifically, building upon the approach of AppAgent \cite{appagent}, the agent is tasked with autonomously exploring a specific app to collect trajectories. However, different from AppAgent, which is inefficient and focuses on specific UI elements, we concentrate on the functionality of the entire page. For each page at step $t$ in a trajectory, given the observation $o_t$ and the historical actions $[a_0, a_1, \dots, a_{t-1}]$, we prompt a multimodal LLM to analyze and summarize the following information:
\begin{itemize}
  \item
    \textbf{Page Description:} It refers to a description of the content, structure, and functions of an application page.
  \item
   \textbf{Key UI Elements:} The functional UI elements in the page that play a ``key role" in user interaction, page navigation, or page jumps. Including but not limited to Toolbar, Menu, Label, Button, ImageView, BottomNavigationView.
  \item
  \textbf{Action Path:} A summary description of the path from the initial page to the current page.
  \item
  \textbf{Page Label:} A concise and descriptive page label generated based on the above understanding of the page description and key UI elements.
\end{itemize}
This information is constructed as a structured page chunk. It is vectorized and stored in a collection within the vector database, forming long-term memory. Each collection holds page memory for a specific app, which aims to prevent confusion from similar basic elements (e.g., search bars, appearance settings) across different apps.


\subsection{Memory-Augmented Task Planning}
When faced with a user task, the agent plans the task with the aid of the memory database, establishing an organized and systematic foundation for task execution.
To effectively utilize the memory database, we introduce a coarse-to-fine task planning approach.
Given a user task $T$, the agent first performs a coarse-grained planning phase, during which the LLM Planner generates coarse-grained subtasks:
\begin{equation}
    \mathcal{T}_{cg} = \left \{  s_1, s_2, \dots, s_k \right \} = PL(T)
\end{equation}
where $PL$ represents the LLM Planner, each $s_k \in \mathcal{T}_{cg}$ represents a coarse-grained subtask. Subsequently, these subtasks are assigned to the appropriate apps by an LLM Task Scheduler, ensuring that each subtask can be executed in the corresponding app. The following formula represents this process:
\begin{equation}
\begin{split}
    \mathcal{S}_{app} 
    &= \left \{S_{app_1}, S_{app_2}, \dots, S_{app_n} \right \} \\
    &= \left \{(app_1, s_1), (app_2, s_2, s_3), \dots (app_n,s_m)\right \} \\
    &= TS(\mathcal{T}_{cg}, APPs)
\end{split}
\end{equation}
where $TS$ represents the LLM Task Scheduler, $APPs$ represents all installed applications, $app_i$ refers to a specific app, $S_{app_n}$ denotes the subtasks in $app_n$. 
Then, in the fine-grained planning phase, subtasks in each app serve as queries to retrieve top-$k$ ($k$=3) relevant memory pages from the corresponding app:
\begin{equation}
\mathcal{P}_{app_i} = \left \{ v_j \in D_i | \mathrm {rank} (v_j, Emb(S_{app_i})) < k \right \}
\end{equation}
where $D_i$ represents the collection corresponding to $app_i$, $v_j \in D_i$ is the stored page vector, $Emb()$ is the embedding function, and $\mathrm {rank}$ represents the ranking of the vector $v_j$ under query $S_{app_i}$, based on cosine similarity.
The retrieved pages $\mathcal{P}_{app_i}$ serve as supplementary context to further instruct the LLM Planner to generate a refined secondary plan, which is formally represented by the following equation:
\begin{equation}
\mathcal{T}_{fg} = \left \{PL(S_{app_i}, \mathcal{P}_{app_i} ) | i=1,2, \dots, \left | \mathcal{S}_{app} \right | \right \}
\end{equation}
where $\mathcal{T}_{fg}$ is a fine-grained list of subtasks that is regenerated for each app. With the injected memory pages, the agent can understand which UI elements of the corresponding app need to be used to complete the task and take the corresponding arrival path as an example to guide planning.


\subsection{Task Executor}
The task executor, which acts as a base agent, is responsible for directly interacting with the mobile device environment. Following the planning phase, the task executor iteratively executes the planned tasks and determines the specific operations to be performed. 

In real-world dynamic digital environments,  LLM struggles to detect and correct simple errors and is prone to failure due to inadequate evaluation of task progress. However, it has been demonstrated that LLM can effectively serve as a ``reviewer" by evaluating the outputs of other models \cite{Zheng23LLMasajudge, li2025generationjudgment}. Motivated by this, we developed a dual-LLM architecture executor comprising two roles, a Decision-maker and a Judge, both supported by a multimodal LLM that collaborate to execute tasks. At the initial state, the Decision-maker determines the first action $a_{0}$ based on the fine-grained task $\mathcal{T}_{fg}$, the current SoM \cite{yang2023setofmark} visual observation $\tilde{o}_{0}$. And we adopt ReAct-style \cite{yao2023react} action generation, which integrates reasoning thought before each action. This process is formalized as follows:
\begin{equation}
    (\Theta_0, a_0) = DM(\mathcal{T}_{fg}, \tilde{o}_{0})
\end{equation}
where $\Theta_0$ denotes the reasoning thought behind action $a_0$, and $DM$ is the LLM Decision-maker.
After the action is executed, the Judge evaluates the progress of the task by analyzing the changes in observations before and after the action, along with the previous reasoning thought. 
It generates a reasoning chain to evaluate  whether the prior action is successful, the current task progress following the prior operation, and recommendations for the next action. For each subsequent time step $t$, this process is formulated as:
\begin{equation}
    (\mathtt{E},\mathtt{P},\mathtt{A^{+}})_t = JU(\mathcal{T}_{fg}, o_{t-1}, o_t, \Theta_{t-1}, \mathtt{P}_{t-1})
\end{equation}
where $\mathtt{E}$ indicates the evaluation for the previous action, $\mathtt{P}$ represents the current task progress, $\mathtt{A^{+}}$ is the suggestion for next step, all of which are expressed as natural language strings. And $JU$ is the LLM Judge. By integrating the evaluation results from the Judge, the Decision-maker proceeds to make the next decision according to the following formulation:
\begin{equation}
    (\Theta_t, a_t) = DM(\mathcal{T}_{fg}, \tilde{o}_{t}, (\mathtt{E},\mathtt{P},\mathtt{A^{+}})_t)
\end{equation}

We use a short-term memory unit to maintain the historical responses generated at each step of the current task. Furthermore, considering that certain operations may require task-related information from previous screens in complex tasks, we prompt LLM to record task-related information based on the relationships between subtasks. This provides necessary information support when performing specific page operations, such as cross-platform comparison of product prices.


\section{Experiments}
To evaluate the effectiveness of our method in performing GUI interactions on mobile devices, we conduct experiments on a diverse set of widely used apps in real-world scenarios.

\subsection{Experiment Setup}
\paragraph{Implementation Details}
We carry out all experiments on a Xiaomi 14 and use Android Debug Bridge (ADB) as the tool to execute operations. For the agent model, all experiments are conducted with GPT-4o \cite{openai2024gpt4ocard} to ensure consistent model capabilities across tests.  For the vector database, we use Milvus \cite{Milvus} to store trajectory memory. For the embedding model, we access text-embedding-v3 from Alibaba Cloud.

\paragraph{Benchmark}
We evaluate our method using two real-life scenario test sets: SPA-Bench \cite{chen2025spabench} and the CHOP dataset \cite{zhou2025chop}. SPA-Bench consists of 340 tasks, including 150 single-app tasks and 20 cross-app tasks, across both English and Chinese apps, involving 58 third-party apps. The CHOP dataset comprises two subsets representing English and Chinese application scenarios. Each subset contains 10 applications, with three tasks per application categorized by difficulty level: easy, medium, and hard, for a total of 60 tasks. The tasks in both test sets are designed with progressively increasing complexity while emphasizing core app functionalities that reflect daily usage scenarios, thereby offering a realistic evaluation of smartphone agent performance. More details about the benchmarks are provided in Appendix B.

\begin{table}[!bp]
\scriptsize
\begin{tabular}{@{}lcccc@{}}
\toprule
\multirow{2}{*}{Agent} & \multicolumn{2}{c}{Single-App} & \multicolumn{2}{c}{Cross-App} \\  
\cmidrule(lr){2-3} \cmidrule(l){4-5}
& English        & Chinese       & English       & Chinese       \\   
\midrule
AutoDroid \cite{AutoDroid2024}    & 0.327          & 0.187         & 0            & 0  \\
AppAgent \cite{appagent}     & 0.340          & 0.247         & 0            & 0  \\
MobileAgentV2 \cite{mobileagentv2} & 0.433       & 0.440      & 0.100         & \underline{0.100}  \\
SeeAct \cite{zheng2024seeact}       & 0.393          & 0.327        & 0.100       & 0.050 \\ 
M3A \cite{rawles2025androidworld}          & \textbf{0.640} & \underline{0.447} & \underline{0.200}  & \underline{0.100} \\
\midrule
Ours          & \underline{0.627}      & \textbf{0.553}    & \textbf{0.350}    & \textbf{0.350}        \\ 
\bottomrule
\end{tabular}
\centering
\caption{Performance comparison across all scenarios in SPA-Bench. The best and second-best performances are in bold and underlined, respectively. 
}
\label{tab:spa-bench}
\end{table}

\begin{table*}[t]
\small
\setlength{\tabcolsep}{1mm}
\begin{tabular}{lcccccccc}
\toprule
\multirow{2}{*}{Agent} & \multicolumn{4}{c}{English}    & \multicolumn{4}{c}{Chinese}    \\
\cmidrule(lr){2-5} \cmidrule(lr){6-9}
                       & Easy & Medium & Hard & Overall & Easy & Medium & Hard & Overall \\
\midrule
AppAgent \cite{appagent}               & 0.5  & 0.4    & 0.1  & 0.333   & 0.4  & 0.3    & 0.2  & 0.300   \\
MobileAgentV2 \cite{mobileagentv2}         & 0.5  & 0.5    & 0.4  & 0.467   & \underline{0.8}  & 0.2    & \underline{0.3}  & 0.433   \\
MOBA \cite{zhu2025moba}                  & 0.5  & 0.3    & 0.2  & 0.333   & 0.4  & 0.3    & 0.1  & 0.267   \\
CHOP \cite{zhou2025chop}                  & \underline{0.8}  & \underline{0.7}    & \textbf{0.6}  & \underline{0.700}   & \textbf{1.0 } & \textbf{0.8}    & 0.1  & \underline{0.633}   \\
\midrule
Ours                   & \textbf{0.9}  & \textbf{0.9}    & \textbf{0.6}  & \textbf{0.800}   & \underline{0.8}  & \textbf{0.8}    & \textbf{0.5}  & \textbf{0.700}   \\
\bottomrule
\end{tabular}
\centering
\caption{Performance comparison on English and Chinese tasks, categorized by difficulty in the CHOP dataset.}
\label{tab:chop}
\end{table*}

\begin{figure*}
  \centering
  \includegraphics[width=0.8\textwidth]{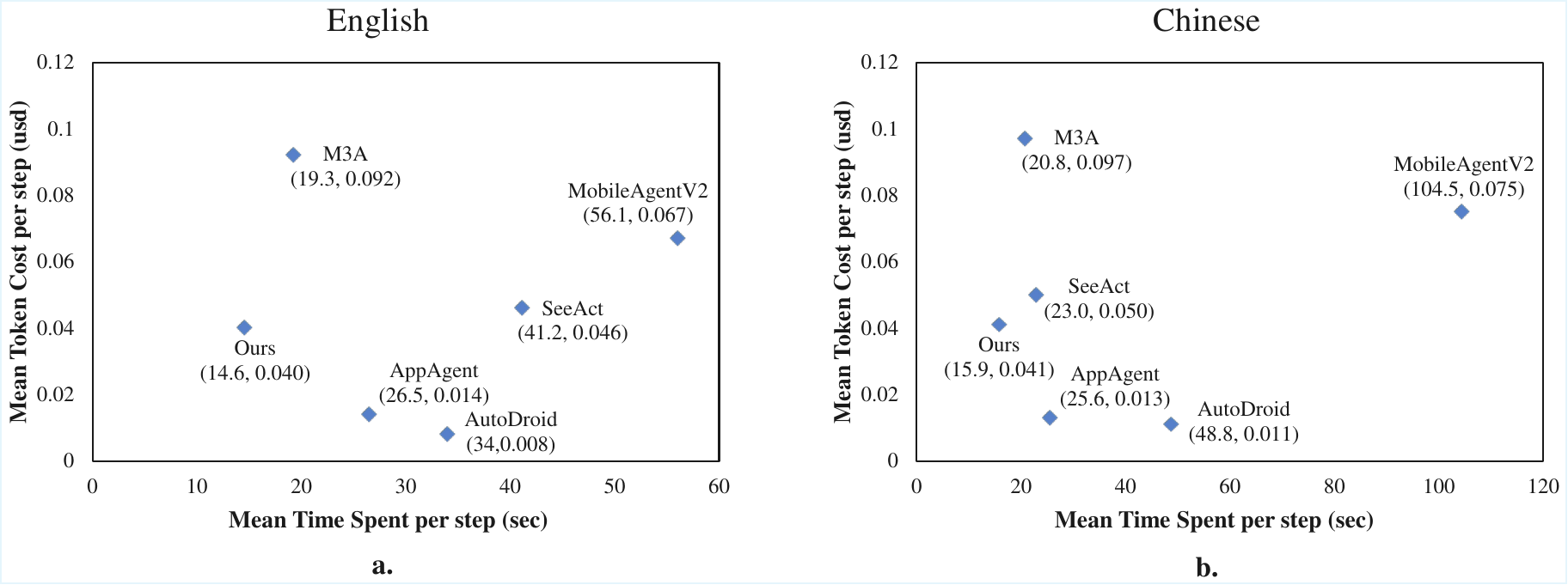}
  \caption{Computational overhead analysis. Mean Time Spent per step (MTS)$\downarrow$: Evaluates the agent's operational response speed, defined as the average time (in seconds) consumed by the agent to process and execute each action step. Mean Token Cost per step (MTC)$\downarrow$: Assesses the economic efficiency of agent operation, representing the average monetary cost (in US dollars) attributable to LLM API usage for generating each action step.}
  \label{fig: Overhead Analysis}
\end{figure*}

\subsection{Experimental Results}
\paragraph{Success Rate Comparison}
We compare with five leading agents in SPA-Bench, including: AutoDroid \cite{AutoDroid2024}, AppAgent \cite{appagent}, MobileAgentV2 \cite{mobileagentv2}, SeeAct \cite{zheng2024seeact}, M3A \cite{rawles2025androidworld}.
\textbf{Table \ref{tab:spa-bench} shows the success rate results of using the GPT-4o backbone for all agents in SPA-Bench.} For single-app tasks, compared to M3A, MapAgent shows an improvement of 10.6\% in Chinese scenarios while maintaining comparable performance in English scenarios. For more challenging cross-app tasks, MapAgent shows notable improvement in both English and Chinese settings. This is primarily due to our coarse-to-fine planning approach, which decomposes a cross-app task into a sequence of single-app subtasks in the coarse-grained planning phase. Hence, the agent's context window remains within a reasonable range. And short-term memory helps establish the connection between these subtasks. 
We further compare several agent systems featured in the CHOP dataset, including: AppAgent, MobileAgentV2, MOBA \cite{zhu2025moba}, CHOP \cite{zhou2025chop}.
\textbf{Table \ref{tab:chop} presents the success rate results in the CHOP dataset.} In both English and Chinese tasks, our method achieves the highest overall success rate as well as elevated success rates in both easy and medium difficulty tasks. In hard difficulty tasks, although there is a slight decrease, it still outperforms other methods. These results demonstrate the effectiveness and adaptability of our method.



\paragraph{Computational Overhead Analysis}
For computational overhead analysis, we examine the single-app scenario of SPA-Bench, which encompasses a larger task volume. The results are presented in Figure \ref{fig: Overhead Analysis}. 
Regarding the response speed for mean time spent per step, our method keeps practical time consumption (15.3s per step average).
When considering mean token cost per step, AutoDroid \cite{AutoDroid2024} exhibits the lowest costs due to its dependence solely on plain text.
In contrast to other methods, our approach achieves a comparatively high success rate while incurring a relatively low cost. These findings demonstrate that our method not only maintains an appreciable success rate but also ensures rapid response times and cost-effectiveness. This indicates a reasonable trade-off between efficiency and effectiveness in our approach.

\begin{figure}
  \centering
  \includegraphics[width=0.45\textwidth, keepaspectratio]{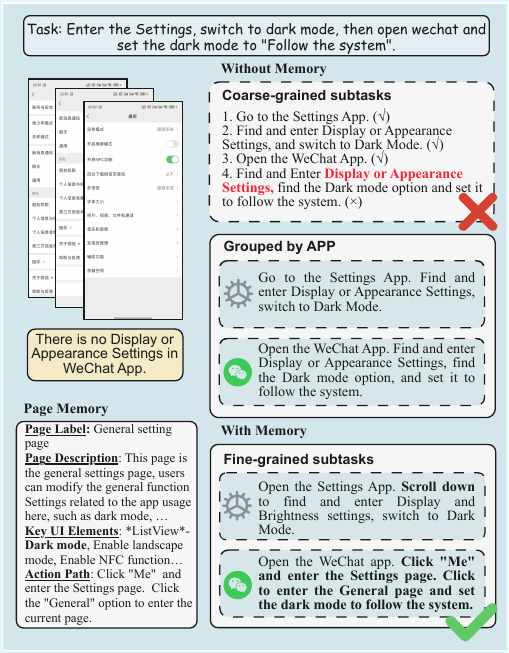}
  \caption{An example of a cross-app task planning process.}
  \label{fig: case study}
\end{figure}

\begin{table}[!b]
\small
\begin{tabular}{lccc}
\toprule
Configuration & SR  & MTC & MTS \\ 
\midrule
\multicolumn{4}{c}{English}                  \\
\midrule
MapAgent \textit{w/o M \& J} & 0.428 & \textbf{8.6}  & \textbf{0.021} \\
MapAgent \textit{w/o M}      & 0.531 & 14.1 & 0.038 \\
MapAgent              & \textbf{0.627} & 14.6 & 0.040  \\
\midrule
\multicolumn{4}{c}{Chinese}                  \\
\midrule
MapAgent \textit{w/o M \& J} & 0.320  & \textbf{9.8}  & \textbf{0.023} \\
MapAgent \textit{w/o M}      & 0.433 & 15.1 & 0.039 \\
MapAgent              & \textbf{0.553} & 15.9 & 0.041 \\
\bottomrule
\end{tabular}
\centering
\caption{Ablation Study on single-app tasks in SPA-Bench. While success rate gradually increases,  this trend reveals the practical trade-offs. Each response generated by Judge (\textit{J}) necessitates an additional round of access to the LLM API.
Employing memory (\textit{M}) to augment planning causes additional costs of retrieval and a round of secondary planning.}
\label{tab:ablation study}
\end{table}


\subsection{Ablation Study}
To explore the impact of the proposed memory-augmented planning mechanism and dual-LLM architecture, we conduct comprehensive ablation studies. The experimental results, which utilize various configurations to assess agent behavior in the single-app task scenario within SPA-Bench, are presented in Table \ref{tab:ablation study}. It reveals practical tradeoffs. When both the memory for planning and the Judge for evaluation are disabled (\textit{w/o M \& J}), the success rate is the lowest. However, this configuration also leads to minimal time and cost consumption. Conversely, when only memory is disabled (\textit{w/o M}), there is a corresponding increase in success rate; however, this improvement comes with a notable increase in time and cost consumption. This observation is justifiable, as each Judge's response necessitates an additional round of access to the LLM API. When memory is also employed for planning augmentation, there continues to be an upward trend in success rate while only experiencing a slight increase in time and cost consumption, primarily attributed to delays from the retrieval of page memory and the added step for fine-grained planning before task execution.

To further elucidate the roles of the memory mechanism and the Judge, we analyze variations in success rates across different task levels within the Chinese single-app scenario, as illustrated in Table \ref{tab:different levels}. The introduction of Judge yields significant performance improvement at Levels 2 and 3. 
These complex tasks typically require more steps to complete, highlighting both the effectiveness and necessity of comprehension for task progress. 
With memory for augmenting planning, we observe improvements in success rates across all task levels. To explain this, it is important to note that in the fine planning stage, not only are the key UI elements in page retrieved from memory used to compensate for agent's insufficient understanding of a certain application, but the action path also serves as an exemplar, thereby effectively functioning as a few-shot learning mechanism \cite{llmfewshot, min2022rethinkingrole}.

\begin{table}
\small
\begin{tabular}{lccc}
\toprule
Configuration           & Level 1 & Level 2 & Level 3 \\ \midrule
MapAgent \textit{w/o M \& J}   & 0.64    & 0.24    & 0.08    \\
MapAgent \textit{w/o M}        & 0.68    & 0.40     & 0.22    \\
MapAgent              & \textbf{0.76}    & \textbf{0.56}    & \textbf{0.34}    \\
\bottomrule
\end{tabular}
\centering
\caption{Success rates of different difficulty levels on single-app Chinese tasks in SPA-Bench.}
\label{tab:different levels}
\end{table}

\begin{table}
\small
\begin{tabular}{lc}
\toprule
Error Type                      & \multicolumn{1}{l}{Proportion} \\ \midrule
Poor UI Recognition             & 36\%                           \\
Misunderstandings of task and context & 25\%   \\
XML Output Error                & 15\%                           \\
Step Omission                   & 12\%                           \\
Others                          & 12\%                           \\ \bottomrule
\end{tabular}
 \centering
 \caption{Error distribution.}
 \label{tab: error analysis}
\end{table}

\subsection{Case Study and Error Analysis}
Figure \ref{fig: case study} illustrates the planning process of a cross-app task. Without prior knowledge of page memory, the agent correctly planned the task segment in Settings app but failed to configure dark mode in WeChat app. After subtasks were scheduled to the appropriate apps and the corresponding pages were retrieved for in-context planning, the error in the initial coarse planning was resolved, and explicit path instructions were provided. This facilitates decision-making and results in smoother task execution.
The complete execution process and more cases are available in Appendix E.

To promote future work, we perform a detailed analysis of failed tasks in SPA-Bench. We categorize all error types according to the essential causes of each task error and summarize their distribution as shown in Table \ref{tab: error analysis}. 
These errors suggest the direction of our future research.
Further details of error analysis can be found in Appendix C.

\section{Conclusion}
We present MapAgent, an advanced framework for mobile task automation that fundamentally enhances LLM-based planning through trajectory-constructed memory augmentation. By storing historical trajectories and retrieving task-relevant pages via similarity matching, MapAgent substantially mitigates the knowledge bias between LLM and real-world applications. The integration of a specialized dual-LLM architecture task executor ensures stable operation in unpredictable mobile environments through collaborative action generation and real-time progress monitoring. Extensive real-world experiments in both English and Chinese contexts demonstrate the effectiveness of MapAgent, confirming its promising potential for advancing AI-assisted smartphone development.

\bibliography{aaai2026}

\end{document}